\documentclass[preprint]{aastex}
\usepackage{natbib}
\usepackage{color}
\usepackage{ulem}
\slugcomment{LA-UR-11-11505 - \textcolor{red}{9/29/2011 DRAFT for Submission}}

\shorttitle{Observation and Spectral Measurements of the Crab Nebula with Milagro}
\shortauthors{Abdo et al.}

\begin{document}

\title{Observation and Spectral Measurements of the Crab Nebula with Milagro}

\author{
A.~A.~Abdo,\altaffilmark{\ref{msu},\ref{nrl}}
B.~T.~Allen,\altaffilmark{\ref{uci},\ref{cfa}} 
T.~Aune,\altaffilmark{\ref{ucsc}}
W.~Benbow,\altaffilmark{\ref{hsc}} 
D.~Berley,\altaffilmark{\ref{umcp}} 
C.~Chen,\altaffilmark{\ref{uci}}
G.~E.~Christopher,\altaffilmark{\ref{nyu}}
T.~DeYoung,\altaffilmark{\ref{psu}} 
B.~L.~Dingus,\altaffilmark{\ref{lanl}} 
R.~W.~Ellsworth,\altaffilmark{\ref{georgemason}} 
A.~Falcone,\altaffilmark{\ref{psu}} 
L.~Fleysher,\altaffilmark{\ref{nyu}} 
R.~Fleysher,\altaffilmark{\ref{nyu}} 
M.~M.~Gonzalez,\altaffilmark{\ref{ida}} 
J.~A.~Goodman,\altaffilmark{\ref{umcp}}
J.~B.~Gordo,\altaffilmark{\ref{udm}} 
E.~Hays,\altaffilmark{\ref{gsfc}},
C.~M.~Hoffman,\altaffilmark{\ref{lanl}}
P.~H.~H\"untemeyer,\altaffilmark{\ref{utah}}
B.~E.~Kolterman,\altaffilmark{\ref{nyu}} 
J.~T.~Linnemann,\altaffilmark{\ref{msu}}
J.~E.~McEnery,\altaffilmark{\ref{gsfc}}
T.~Morgan,\altaffilmark{\ref{unh}}
A.~I.~Mincer,\altaffilmark{\ref{nyu}} 
P.~Nemethy,\altaffilmark{\ref{nyu}} 
J.~Pretz,\altaffilmark{\ref{lanl}}
J.~M.~Ryan,\altaffilmark{\ref{unh}} 
P.~M.~Saz~Parkinson,\altaffilmark{\ref{ucsc}}
A.~Shoup,\altaffilmark{\ref{osu}} 
G.~Sinnis,\altaffilmark{\ref{lanl}} 
A.~J.~Smith,\altaffilmark{\ref{umcp}} 
V.~Vasileiou,\altaffilmark{\ref{umcp},\ref{cresst}} 
G.~P.~Walker,\altaffilmark{\ref{lanl},\ref{nst}} 
D.~A.~Williams\altaffilmark{\ref{ucsc}}
and 
G.~B.~Yodh\altaffilmark{\ref{uci}}} 

\altaffiltext{1}{\label{msu} Department of Physics and Astronomy, Michigan State University, 3245 BioMedical Physical Sciences Building, East Lansing, MI 48824}
\altaffiltext{2}{\label{nrl} Current address: Space Science Division, Naval Research Laboratory, Washington, DC 20375}
\altaffiltext{3}{\label{uci} Department of Physics and Astronomy, University of California, Irvine, CA 92697}
\altaffiltext{4}{\label{cfa} Current address: Harvard-Smithsonian Center for Astrophysics, Cambridge, MA 02138}
\altaffiltext{5}{\label{ucsc} Santa Cruz Institute for Particle Physics, University of California, 1156 High Street, Santa Cruz, CA 95064}
\altaffiltext{6}{\label{umcp} Department of Physics, University of Maryland, College Park, MD 20742}
\altaffiltext{7}{\label{nyu} Department of Physics, New York University, 4 Washington Place, New York, NY 10003}
\altaffiltext{8}{\label{psu} Department of Physics, Pennsylvania State University, University Park, PA 16802}
\altaffiltext{9}{\label{lanl} Group P-23, Los Alamos National Laboratory, P.O. Box 1663, Los Alamos, NM 87545}
\altaffiltext{10}{\label{georgemason} Department of Physics and Astronomy, George Mason University, 4400 University Drive, Fairfax, VA 22030}
\altaffiltext{11}{\label{ida} Instituto de Astronom\'ia, Universidad Nacional Aut\'onoma de M\'exico,
D.F., M\'exico, 04510}
\altaffiltext{12}{\label{gsfc} NASA Goddard Space Flight Center, Greenbelt, MD 20771}
\altaffiltext{13}{\label{utah} Department of Physics, University of Utah, Salt Lake City, UT 84112}
\altaffiltext{14}{\label{unh} Department of Physics, University of New Hampshire, Morse Hall, Durham, NH 03824} 
\altaffiltext{15}{\label{osu} Ohio State University, Lima, OH 45804}
\altaffiltext{16}{\label{cresst} CRESST NASA/Goddard Space Flight Center, MD 20771 and University of Maryland, Baltimore County, MD 21250}
\altaffiltext{17}{\label{nst} Current address: National Security Technologies, Las Vegas, NV 89102}
\altaffiltext{18}{\label{hsc}Harvard-Smithsonian Center for Astrophysics, Cambridge, MA 02138}
\altaffiltext{19}{\label{udm}Departmento de F\'sica, Universidad de Murcia, E-30100 Murcia, Spain}

\begin{abstract}

The Crab Nebula was detected with the Milagro
experiment at a statistical significance of 17 standard deviations 
over the lifetime of the experiment.
The experiment was sensitive to approximately 100 GeV - 100 TeV gamma ray air showers
by observing the particle footprint reaching the ground. 
The fraction of
detectors recording signals from photons at the ground is a suitable proxy
for the energy of the primary particle and has been used to measure the
photon energy spectrum of the Crab Nebula between $\sim$1 and $\sim$100 TeV. 
The TeV emission is believed
to be caused 
by inverse-Compton up-scattering scattering of ambient photons by 
an energetic electron population. The location of a TeV steepening or 
cutoff in the energy spectrum reveals important details about the underlying
electron population.
We describe the experiment and the technique for
distinguishing gamma-ray events from the much more-abundant
hadronic events. We describe the 
calculation of the significance of the excess from the Crab and how 
the energy spectrum is fit.
The excess from the Crab is fit to a function of the form

\begin{center}
$\frac{dN}{dE}(I_0,\alpha,E_{\rm cut}) = I_0 (\frac{E}{E_{0}})^{-\alpha} \exp(\frac{-E}{E_{\rm cut}})$ 
\end{center}

where the flux $I_{0}$, spectral index $\alpha$ and
exponential cutoff energy $E_{\rm cut}$ are allowed to vary and 
$E_{0}$ is chosen to de-correlate the fit parameters. 
The energy spectrum, including the statistical errors from the fit, 
obtained using the simple power law hypothesis, 
that is $E_{cut}=\infty$,
for data between September 2005 and March 2008
is:

\begin{center}
${{dN} \over {dE} }= (6.5\pm0.4) \times 10^{-14} ({E / {10 \:\rm TeV}})^{-3.1\pm 0.1}  {( \rm{cm^2} \: sec \: TeV)}^{-1}$ 
\end{center}

between $\sim$1 TeV and $\sim$100 TeV. When
a finite $E_{cut}$ is fit the result is

\begin{center}
${{dN} \over {dE} }= (2.5^{+0.7}_{-0.4}) \times 10^{-12} ({E / {3 \:\rm TeV}})^{-2.5\pm 0.4} \exp(-E/32^{+39}_{-18} \: \rm{TeV})  {( \rm{cm^2} \: sec \: TeV)}^{-1}$ 
\end{center}

The results are subject to an $\sim$30\% systematic uncertainty in the 
overall flux and an $\sim$0.1 in the power law indicies quoted. Uncertainty
in the overall energy scale has been absorbed into these errors.

Fixing the spectral index to values that have been measured below 1 TeV
by IACT experiments (2.4 to 2.6), 
the fit to the Milagro data suggests that Crab 
exhibits a spectral steepening or cutoff 
between 
about 20 to 40 TeV.

\end{abstract}


\section{Introduction}

The Crab supernova remnant is a luminous nearby VHE gamma-ray source
created by a 1054 CE supernova observed on Earth by Chinese,  Arab
and native American astronomers. 
The optically luminous shell 
is easily visible from Earth.
The Crab Nebula lies 2 kpc from the
Earth and is powered by a 33 ms pulsar that 
injects relativistic electrons into the nebula. 
The central pulsar and its surrounding
nebula are among the most widely studied astronomical objects across
the entire electromagnetic spectrum.
A population of energetic electrons is created by the conversion of rotational
kinetic energy of the neutron star and acceleration in the shock formed
where this flow reaches the surrounding medium. These electrons
can interact by the inverse-Compton process with the associated 
synchrotron photons to create the multi-TeV 
gamma rays that have been seen \citep{gaenslerpwn}.

Despite the recent flares in 100 MeV - 100 GeV 
emission observed in AGILE \citep{agilecrabflare} 
and the Fermi-LAT
\citep{fermiflare} 
and in the TeV by ARGO-YBJ \citep{argoflare}, 
the TeV emission from the Crab is 
believed to be 
steady when observed over several months and
is a standard reference source 
for comparison to other TeV instruments.
Such comparisons are useful as a cross-calibration of the various
ground telescopes.
The Crab was first detected at TeV energies
by the Whipple telescope in 1989 \citep{whipplecrabdiscovery}.
Imaging Atmospheric Cherenkov Telescopes (IACTs)
\citep{hegracrab,hesscrab,veritascrab,magiccrab}
and extensive air-shower (EAS) ground arrays 
\citep{milagrocrab,tibetcrab,argocrab} 
have been used to 
identify and measure the flux of the TeV emission from the Crab.
Since the size of the Crab nebula is small compared to the 
point spread function of TeV gamma-ray detectors,
the emission region appears point-like, and
study of the Crab can serve as a calibration
of an instrument's pointing and angular resolution.

The Milagro experiment \citep{milagrocrab,fermibsl}
was a large water-Cherenkov detector 
sensitive to 
energetic secondary particles in the particle shower resulting when a 
high-energy gamma ray or cosmic ray strikes the atmosphere.  The experiment was able 
to distinguish gamma-ray-induced showers from hadron-induced showers
by measuring the penetrating component characteristic of hadronic particle
showers.
The experiment was sensitive
to EAS resulting from primary gamma rays between 100 GeV and 100 TeV and
has dynamic range to resolve gamma-ray energy spectra between about 1 and
100 TeV.
The experiment operated nearly 24 hours a day
and viewed the entire overhead sky.
The detector was located at 
106.68$^o$W longitude, 35.88$^o$N latitude in northern New Mexico at 
an altitude of 2630~m above sea level and operated from 2000 to 2008.  
The sensitive area of the 
detector comprised two parts: a central water reservoir and 
an ``outrigger'' array of water tanks, both instrumented with 
8-inch hemispherical photomultiplier tubes (PMTs) manufactured by
Hamamatsu (Model R5912). 
The central reservoir was operated alone from 2000 to 2004 when the outrigger
array was added.
The central Milagro reservoir consisted of 
two PMT layers \citep{2000NIMPA.449..478A} 
deployed in a 60x80x8-meter covered water reservoir.  The outrigger array
consisted of 175 water tanks each with a single PMT mounted
at the top of the tank and observing downward into the water of the
Tyvek-lined tank. The outrigger tanks were spread in
an irregular pattern over an area of 200x200 meters around the central 
reservoir. 
The PMTs detected Cherenkov radiation produced in the water by the high-energy
particles in an extensive air shower (EAS) 
that reached to ground level.
The Milagro experiment has previously reported TeV emission from the Crab 
\citep{milagrocrab}
prior to the addition of the Milagro outriggers. With the greater sensitivity
from the outriggers and additional exposure,
measurement of an energy spectrum is possible.
The direction of the primary particle in an EAS
was estimated using the arrival time of the PMT signals.
The angular resolution of Milagro, defined as the standard deviation, $\sigma$,
of a two-dimensional Gaussian fit to the angular error distribution,
varied between about 1.2 and 0.35 degrees for the data presented here.
The angular resolution is a function of both
the size of the 
event, and the operational period of the detector.

Since higher-energy primary particles result in characteristically larger
events on the ground, we use a measure of size of the events on the ground
to measure the spectrum of the Crab, constraining
emission out to 100 TeV. 
In section 
\ref{skymapssection}, we describe the 
background estimation
and the construction of ``skymaps''. 
Section \ref{energyestimationsection} describes the event energy 
estimation and spectral fitting. In section \ref{resultssection}, 
we verify our energy
reconstruction using
cosmic-ray hadrons and finally compute the flux and the 
spectrum of the Crab from 1 to 100 TeV.

\section{Skymaps and Background Estimation}
\label{skymapssection}

From the reconstructed data, a skymap --
a histogram of the sky containing the number of events originating from
each location and associated errors -- is formed.
These skymaps are binned in 
units of 0.1 deg and cover the
viewable sky.
All events
are recorded in the J2000 epoch. 
Each recorded skymap contains a signal 
map and a background map which contain the measured counts on the
sky and the background expectation respectively. 
The skymaps are
constructed in 
independent bins of energy parameter $\mathcal{F}$ which is defined below in 
Equation \ref{frasordef}.

The hadronic background flux is stable in time at TeV energies
because TeV cosmic-rays originate from 
distant sources and propagate diffusively 
in Galactic magnetic fields. Therefore, 
the TeV hadronic 
background is not strongly affected by local variations such as solar activity. 
Instead, the rate and angular distribution of events is dominated by 
variations in 
the atmosphere and the detector.
The background computation technique described below is intended
to measure and correct for these changes. 

\begin{figure}
\plotone{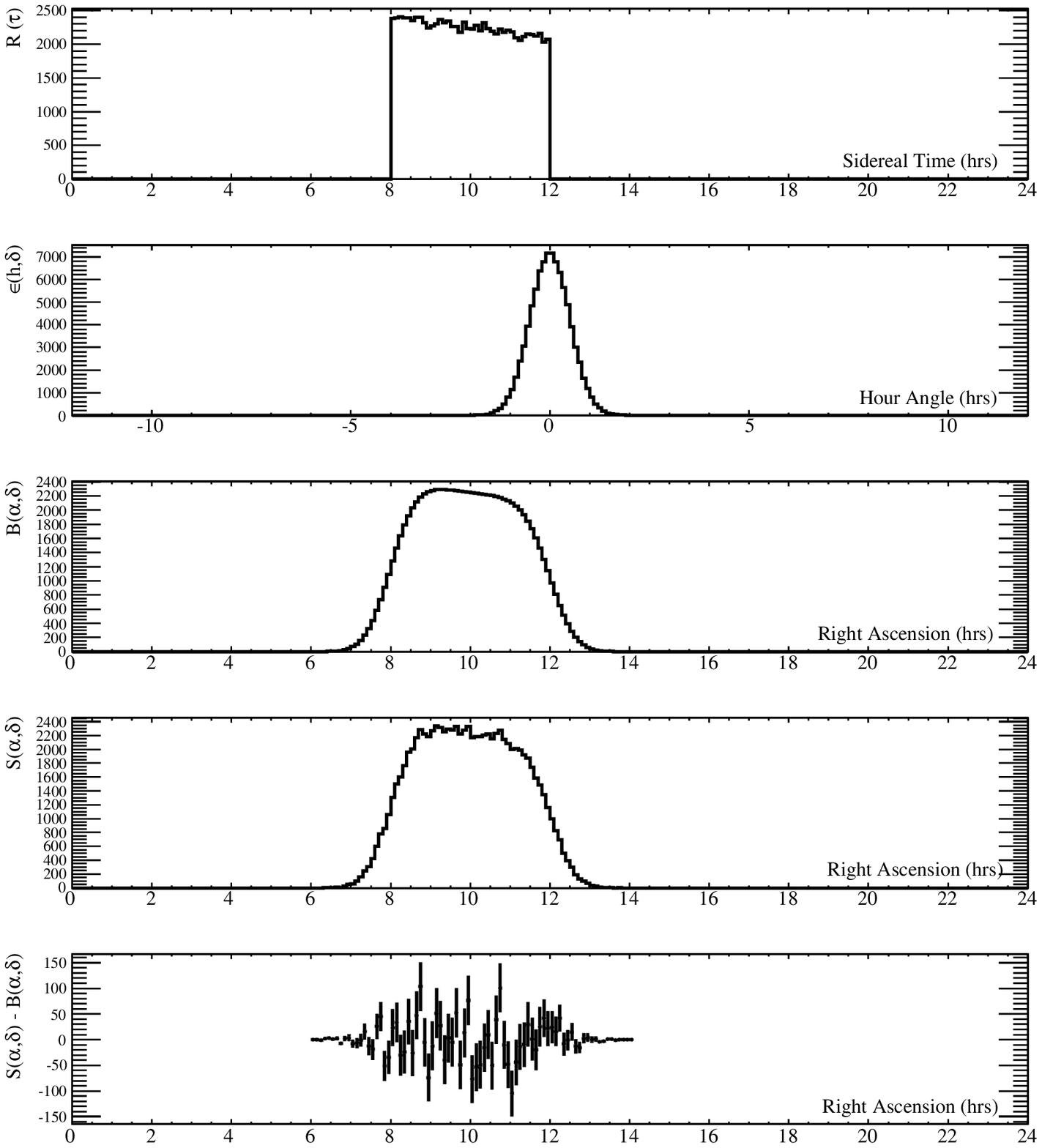}
\caption{\label{twohourintegration}
Example of the background subtraction technique in a single
declination band. For display purposes, this calculation is 
performed with a four-hour integration instead of the standard
two-hour integration. $R(\tau)$ is the all-sky event rate. 
$\epsilon(h,\delta)$ is the local-coordinate distribution of
event arrivals and is convolved with $R(\tau)$ to arrive at
$B(\alpha,\delta)$, the background estimate.  $B(\alpha,\delta)$
is subtracted from the binned event arrival directions $S(\alpha,\delta)$
to arrive at the actual excess estimates. 
}
\end{figure}

The panels of Figure \ref{twohourintegration} demonstrate an example of 
the background
computation in a single declination band.
We represent the background rate $F(\tau,h,\delta)$ 
as a function of sidereal time $\tau$ 
and declination $\delta$ and local hour angle $h$. 
To great
precision, $F(\tau,h,\delta)$ 
can be separated into two independent 
terms, $R(\tau) \cdot \varepsilon(h,\delta)$, where
$R(\tau)$ is the all-sky event rate and
$\varepsilon(h,\delta)$ is the local angular distribution of events.

Even large changes in $R(\tau)$,
due to trigger threshold changes for example, lead to only small
changes in the angular distribution of events on the sky, 
$\varepsilon(h,\delta)$.
We exploit this feature of atmospheric showers to compute the background 
$B(\alpha,\delta)$ in celestial coordinates right-ascension, $\alpha$, and 
declination, $\delta$.  
The technique begins with the definition of an integration
duration.  For most data, the integration duration is two hours but when
looking at rare events (very hard cuts) the integration duration is 
24 hours.  We acquire data for the integration duration and form
the efficiency map $\varepsilon(h,\delta)$.
This efficiency map is a normalized 
probability density function which indicates from where, 
in local detector coordinates,
events arrive.  
The final background estimate for this integration period is the
direct convolution of the efficiency map with the rate.

\begin{equation}
B(\alpha,\delta) = \int \varepsilon(h,\delta) \cdot R(\alpha-h) dh
\end{equation}

We refer to this 
method as ``Direct Integration'' since the local event distribution is 
measured and convolved with the detector's all-sky event rate. The
time-independence of $\varepsilon(h,\delta)$ and the spatial-independence 
of $R(\tau)$ is key because we can use data from the entire integration
duration in the computation of $\varepsilon$ and data from the whole
sky in the calculation of $R$.
This method has been reliably 
demonstrated to estimate backgrounds with systematic errors of
a few parts in $10^{-4}$. The limiting systematic error is
due to real non-uniformities in the cosmic-ray background.

After computing the background estimate $B(\alpha,\delta)$, we can take
the signal map $S(\alpha,\delta)$,
 which is just a histogram of arrival directions and compute the excesses
by computing $S(\alpha,\delta) - B(\alpha,\delta)$ in bins of $\alpha$ and
$\delta$.

\subsection{A5: Gamma/Hadron Separation Parameter}
\label{a5section}

We use an event parameter $A5$ to statistically discriminate air showers
induced by gamma rays from those induced by hadrons.  $A5$ is defined as

\begin{equation}
A5 = \rm{400} \cdot \frac{\mathcal{F} \cdot  \zeta(\rm{t}) \cdot F_{fit} }{MaxPE_{MU}},
\end{equation}

\begin{figure}
\plotone{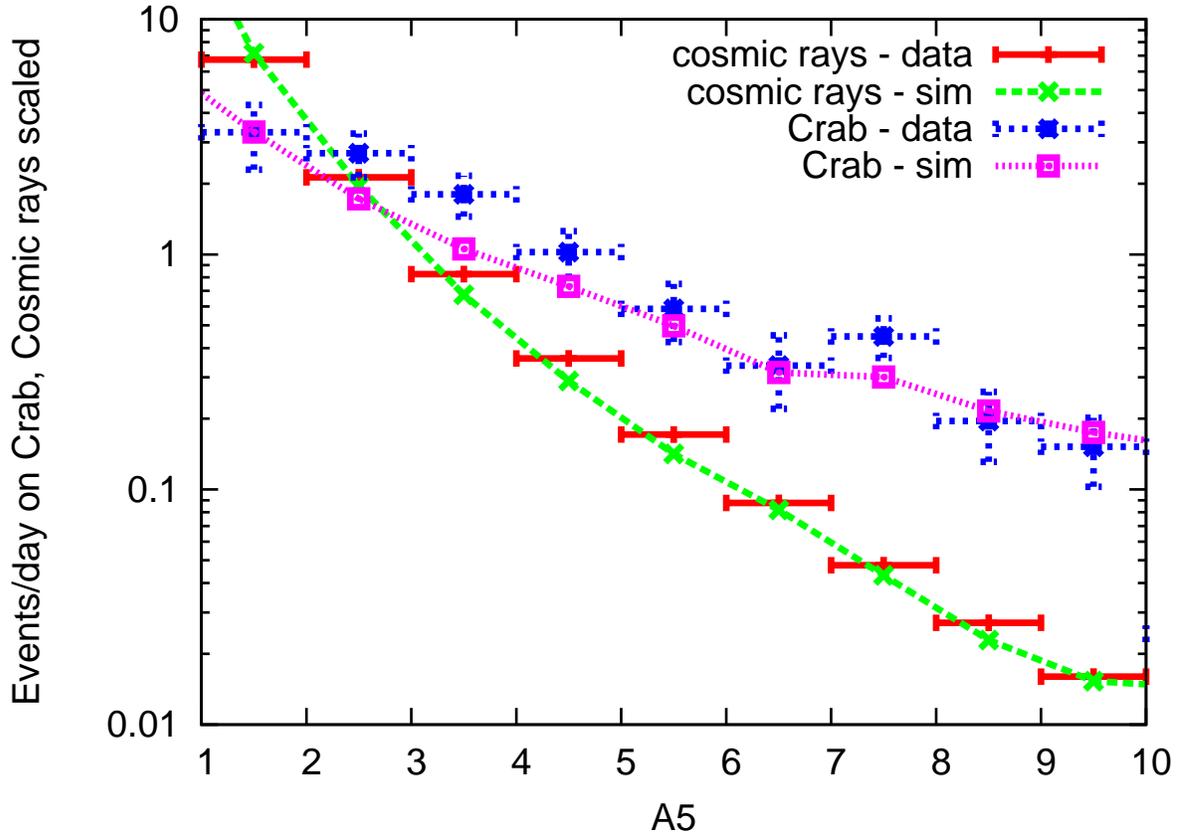}
\caption{\label{a5}
Shown is the distribution of A5 
compared to simulation,
both for the background
cosmic rays and for the background-subtracted Crab excess.
The gamma ray signal is shown 
in a small 0.7 degree circle around the 
true Crab location. The
A5 parameter is used to distinguish gamma-ray events from hadronic events.
A higher value of A5 indicates a higher probability than an event originated
from a gamma ray.
}
\end{figure}

The parameter $\mathcal{F}$ measures 
the size of an event and is defined as

\begin{equation}
\label{frasordef}
\mathcal{F} = {{N_{AS}}\over{N_{AS}^{live}}} + {{N_{OR}}\over{N_{OR}^{live}}},
\end{equation}

where $N_{AS} / N_{AS}^{live}$ is the fraction of live PMTs in the top
layer (or air-shower layer) which participated in the event 
and $N_{OR} / N_{OR}^{live}$ is the fraction of live outriggers to 
participate in the event.  The $\mathcal{F}$ parameter functions is an estimate
of the event's energy and is described more in 
Section \ref{energyestimationsection}.  
The parameter $\rm{F_{fit}}$ is a parameter of the shower-fitting
algorithm indicating what fraction of the PMTs registered times close to the 
fitted shower
plane.
$\rm{MaxPE_{MU}}$
is the number of photo-electrons recorded in the hardest-hit channel
from the bottom layer of the experiment.  The parameter $\zeta(\rm{t})$
is a few percent run-dependent correction to $\rm{F_{fit}}$.
The distribution is seen 
to vary systematically in the data depending on unmodeled factors like
changes
in the calibration. The $\zeta(\rm{t})$ is a correction to take out this
variation in $F_{fit}$.
The $\rm{MaxPE_{MU}}$ in the denominator
of $A5$ is expected to be typically larger for hadron-induced air showers
because the penetrating particles illuminate the bottom layer of the 
experiment.
This means that $A5$ is typically larger for gamma-ray induced showers
with the same number of particles reaching the ground. The
numerator of A5 increases with the size of the event
to account for the fact that we expect more
light in the bottom layer when the event is larger and have to take out the
dependence on the overall size of the event. The overall scaling factor of 400
gives A5 typical values between 1 and 10.
Figure \ref{a5} shows the distribution of A5 for events in a small
circle around the Crab and the separation that A5 provides.

\subsection{Event Weighting}
\label{eventweightingsection}

The $A5$ parameter provides separation
between gamma rays and hadrons primarily because of the higher characteristic
value for gamma-ray sources.  To maximize the statistical significance
when searching for sources, we assign each event a weight based on its
$A5$ value and the signal-to-background expectation for a Crab-like source
for events with that $A5$. A different set of weights is used for each
$\mathcal{F}$ bin.
In this approach, more-gamma-like events
are counted with a higher weight than less-gamma-like events.
A hard cut on the gamma/hadron parameter, which is 
used for the 3 highest $\mathcal{F}$ bins, is simply a step function 
weight.
Weighted skymaps are constructed from data in 9 
$\mathcal{F}$ bins between $\mathcal{F}$ of 0.2 and 2.0 in steps of 0.2.

In addition to the $A5$ weighting for gamma/hadron separation, events are 
given a weight
to
account for the angular resolution of the instrument. 
For a given source position hypothesis, an additional weight is
applied to each event 
which is a function of the angular distance from the 
source position to the reconstructed event position. We assume 
a 2-dimensional Gaussian as the form of the angular resolution
function, where the resolution depends on the event energy parameter, 
$\mathcal{F}$, and ranges from 1.2$^\circ$ for small $\mathcal{F}$ events to 
0.35$^\circ$ for large $\mathcal{F}$ events.


\subsection{Probability Estimation}

In the absence of weighting, events from signal and background 
samples are compared and a probability for the observed 
data under the null hypothesis
can be reliably computed
using Equation 17 of \cite{lima}. When weighting events
rather than simply counting them, we complicate the calculation
of the expected fluctutions.
In the large N limit, this problem has been solved.
If one records not just the sum of the weights, $N = \sum_i w_i$, but 
also the sum of the squares of the weights, $N_2 = \sum_i w_i^2$, 
the error in N is computed as $\delta N = \sqrt(N_2)$. 

\begin{figure}
\plotone{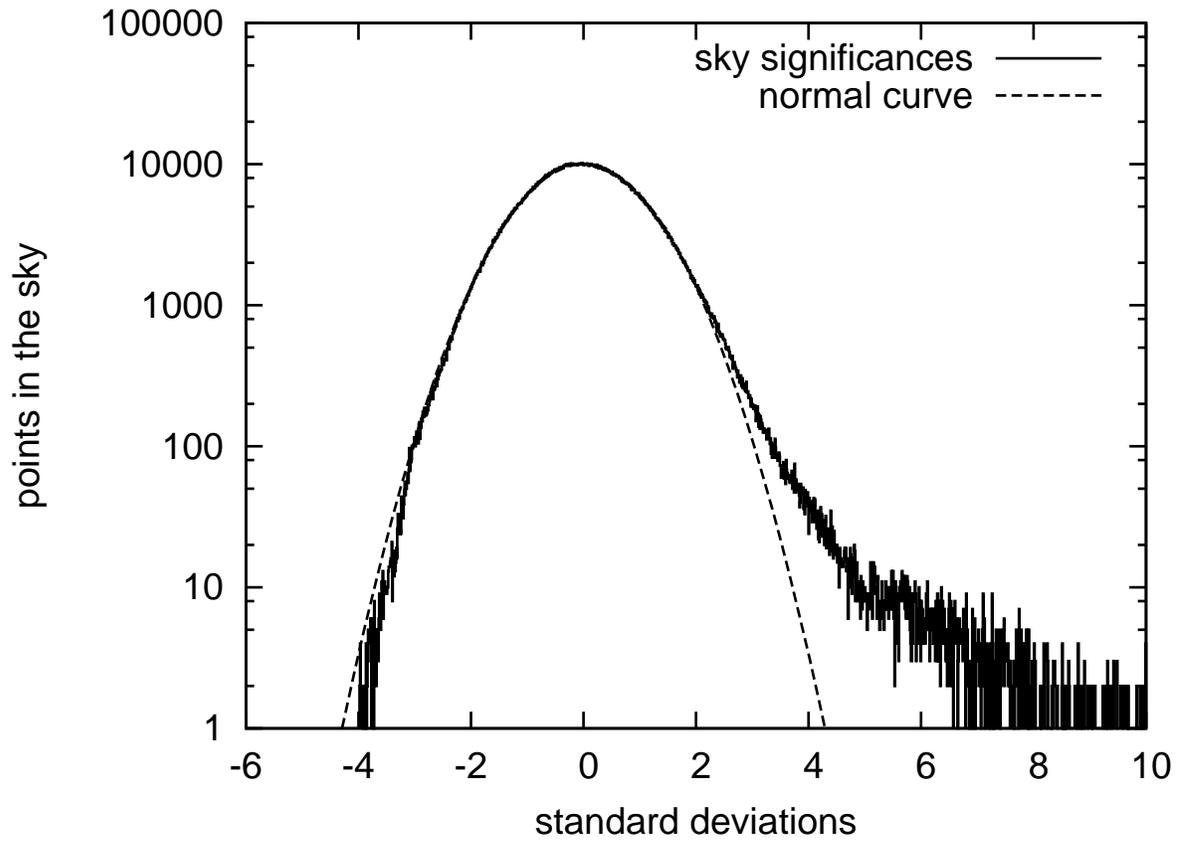}
\caption{\label{sigmadist}
Shown is
the significance distribution of independent points in the Milagro 
field of view. The excess of positive significance is due to the 
presence of sources. The central peak fits well a Gaussian of width
1 $\sigma$, indicating that the statistical significances are calculated
correctly.
}
\end{figure}

\begin{figure}[h]
\plotone{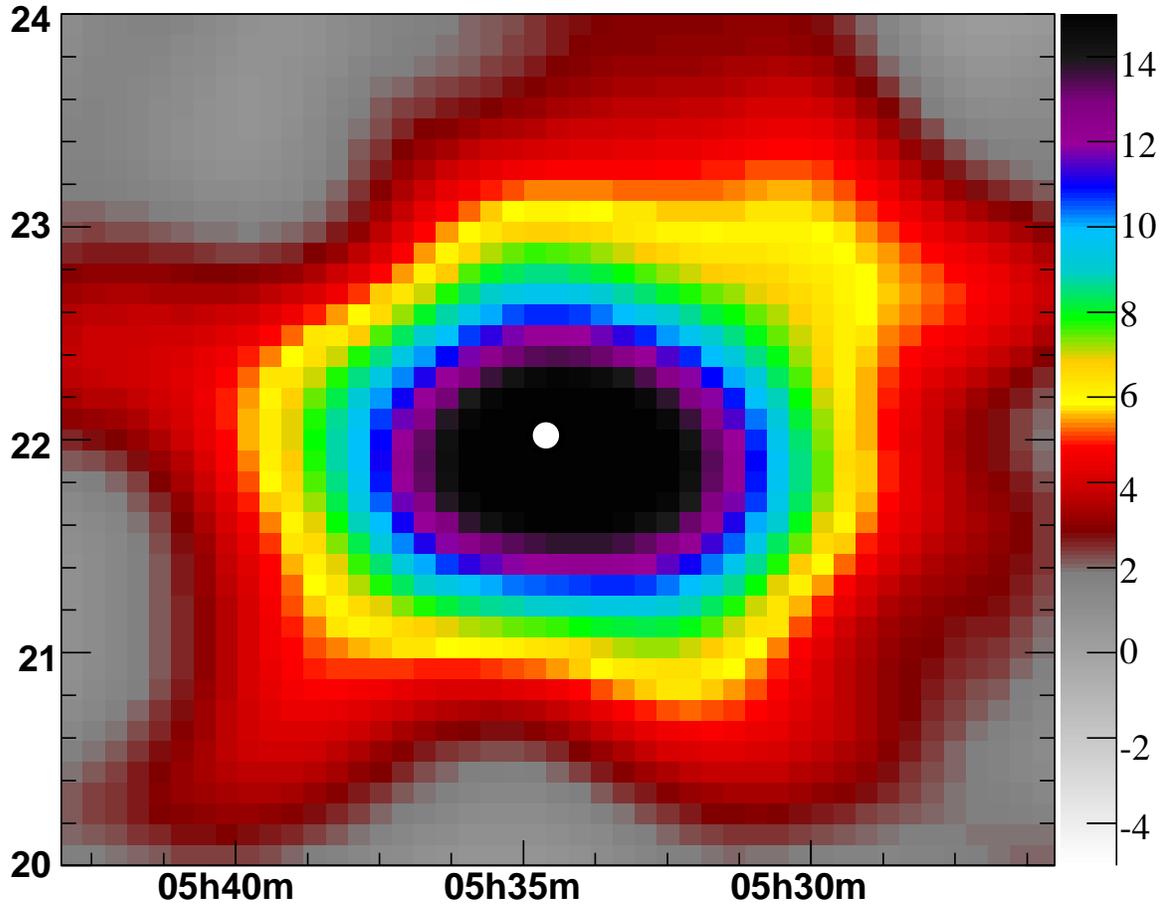}
\caption{\label{crab-image}
Shown is the statistical significance in the region around the Crab Nebula,
indicated by the white dot. The gamma-ray-enhancing weights
as well as the angular smoothing from the text have been 
incorporated. Data over the entire 8-year lifetime of the
experiment have been used and all $\mathcal{F}$ bins have
been combined. At each point in the map, the 
statistical significance is calculated. The smoothing causes the
points to be very correlated. The significance at the location of the Crab is
17.2$\sigma$.
}
\end{figure}

In the small N limit however, the $\delta N = \sqrt(N_2)$ approximation
breaks down, hence the advantage of the approach of Li and Ma who
derived their probability equation assuming Poisson fluctuations.
Poisson distributions take discrete values (integers), unlike continuous
Gaussian distributions. \cite{FeyFeuer} point out that the Poisson 
distributions can be reliably approximated as a continuous envelope 
function described by a single parameter, which for 
a sum of weights is $N^{eff} = \frac{N}{\sqrt{N_2}}$. 
Since fluctuations are well 
approximated as Poisson in $N^{eff}$, we can rewrite the Li and Ma 
expression for significance of an observed result as 
\begin{equation}
S = \sqrt{2}\{ N_{on}^{eff} 
    ln[\frac{1+\alpha}{\alpha} (\frac{N_{on}^{eff}}{N_{on}^{eff}+N_{off}^{eff}})]
   +N_{off}^{eff} 
    ln[(1+\alpha)(\frac{N_{off}^{eff}}{N_{on}^{eff}+N_{off}^{eff}})]\}^{1/2},
\end{equation}
where
\begin{equation}
N_{on}^{eff} = \frac{\sum_i w_{on,i}}{\sqrt{\sum_i w_{on,i}^2}}
\end{equation}
and 
\begin{equation}
N_{off}^{eff} = \frac{\sum_j w_{off,j}}{\sqrt{\sum_j w_{off,j}^2}},
\end{equation}
and $\alpha$ is the usual ratio of the signal and background exposure.
We have studied this approach both through examination of data and 
with Monte Carlo simulations and found it to be reliable 
even in the regime of small statistics, $N^{eff}\sim1$.  Figure \ref{sigmadist}
shows the significance distribution for the Milagro sky. 
Since most of the sky has no gamma ray sources, significances are
distributed normally. 
The fitted mean between $\pm 2 \sigma$
is -0.013$\sigma$ and the width is 0.996$\sigma$.
This is high-level
confirmation that the significance calculation is correct. The 
high-significance tail to this distribution is due to the presence of 
real sources in the sky. Figure \ref{crab-image} shows the final significance
map in the region of the Crab. The significance at the Crab location is 
17.2 standard deviations ($\sigma$). 
This figure includes all data over the 8-year operation of 
Milagro. Only data taken after the outriggers were added are used to 
measure the energy spectrum in this paper.

\section{Energy Estimation}
\label{energyestimationsection}

When a cosmic ray or gamma ray interacts in the atmosphere the amount of 
energy detected at the ground depends on the energy of the 
primary particle and the depth of the initial interaction.
Since the Milagro detector is a large-area calorimeter, it is 
possible to measure the energy reaching the ground level with a relatively small
error ($\sim$20\%). However, fluctuations in the longitudinal development
of air showers - due primarily to fluctuations in the depth of the 
initial interaction - limit the resolution of EAS arrays.
Gamma rays of a given energy that penetrate deeply (a few 
radiation lengths) into the atmosphere deliver substantially more energy 
at the ground level than showers of the same energy
that that interact at the top of the 
atmosphere.
These fluctuations are log-normal \citep{asmithICRC2007} and
dominate the energy resolution for EAS arrays
such as Milagro.
Data from 
September 2005 to March 2008 have been used in determining the 
energy spectrum because the outriggers are needed to provide a dynamic range
spanning 1 to 100 TeV. In this dataset, the statistical significance of the 
Crab is 13.5 standard deviations.


\subsection{The $\mathcal{F}$ Parameter}
\label{frasorsubsection}

\begin{figure}[h] 
  \plotone{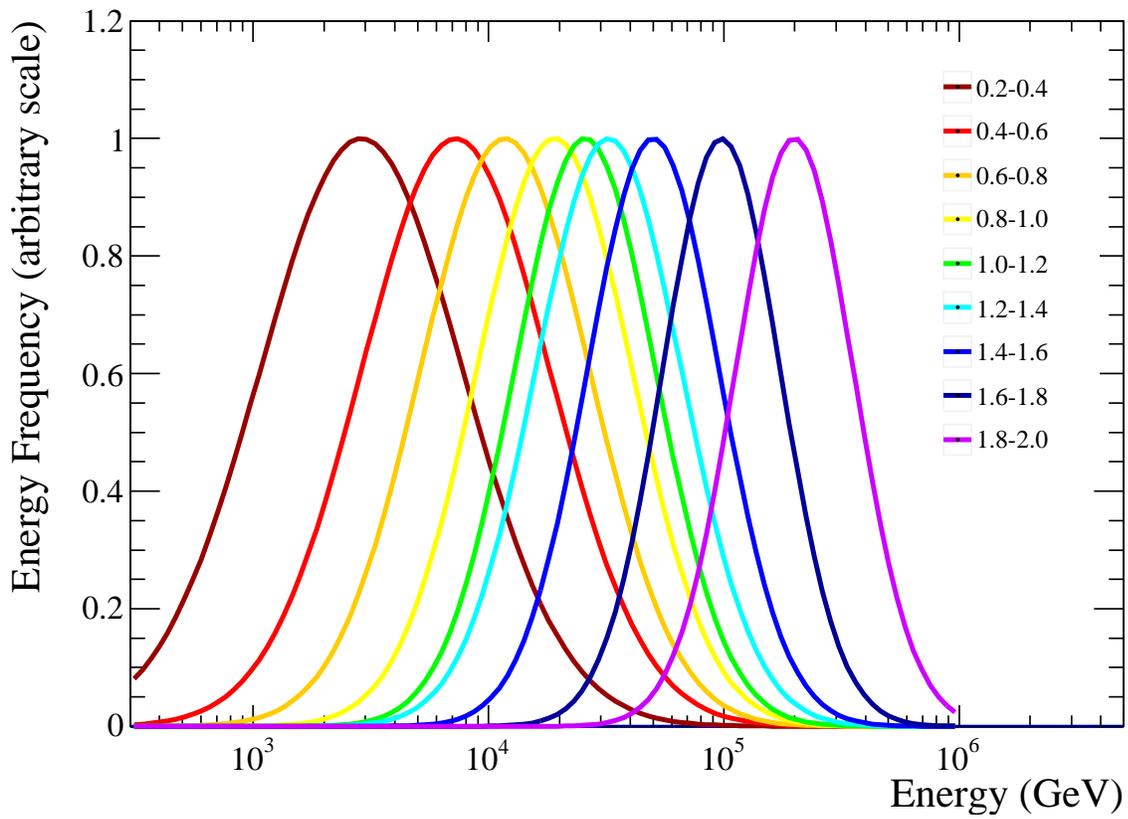}
  \caption{\label{frasorvsenergy} 
    Typical dependence of $\mathcal{F}$ with energy.  
    A source 
    with a spectrum of 2.6 is assumed. Shown is the unit-normalized 
    distribution of true energies for events in the indicated
    $\mathcal{F}$ range.
  } 
\end{figure}

Figure \ref{frasorvsenergy} shows the typical dependence of $\mathcal{F}$,
defined in Equation \ref{frasordef},  
on the 
primary particle energy.  We note that a single $\mathcal{F}$ bin covers a wide
range of energies and that these energies overlap significantly.
Consequently
there is no advantage to a finer segmentation than the 9 bins 
chosen.  

\begin{figure}
  \plotone{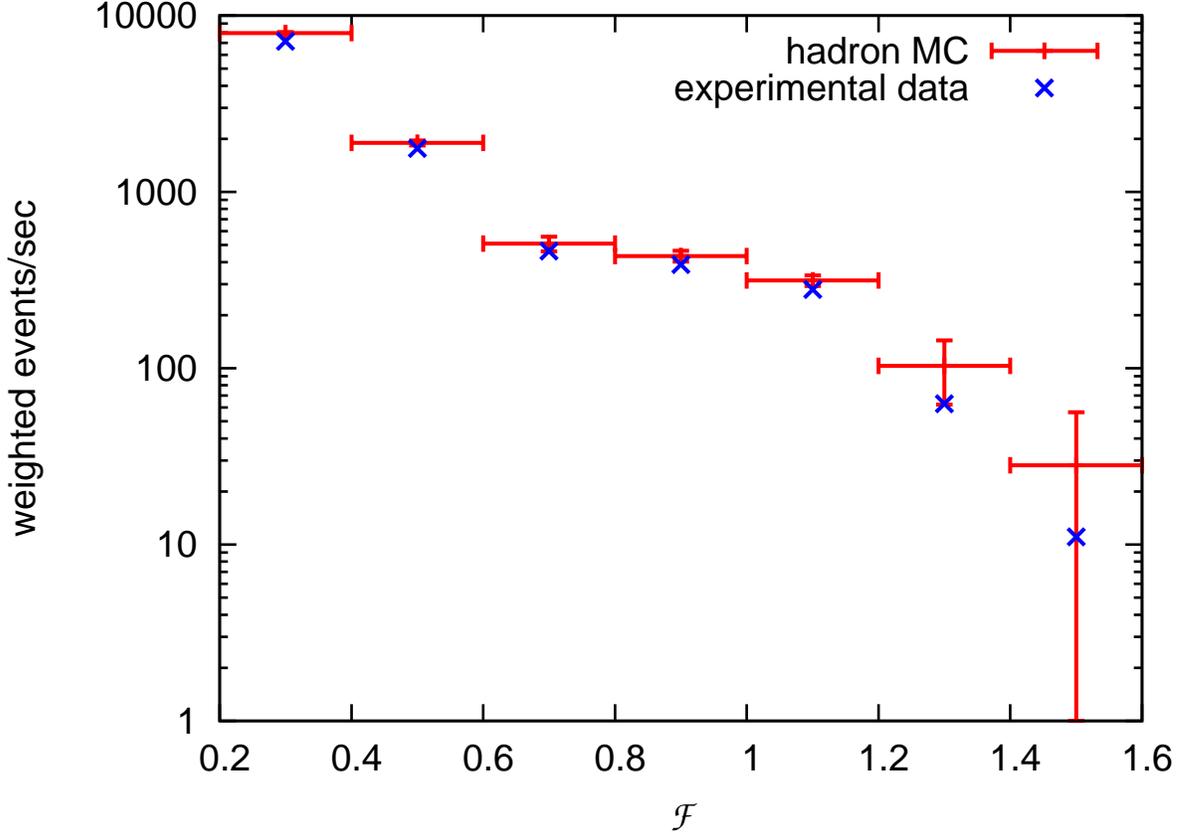}
  \caption{\label{datamcbackground} 
    The $\mathcal{F}$ distribution of the background cosmic rays
    in a 2.5 degree bin at the same declination as the Crab, but 
    offset by a few degrees in right ascension.  Note that
    the weights from Section \ref{eventweightingsection} have been applied in 
    this comparison as a way to probe potential systematic biases introduced by 
    the weighting procedure.  The statistical errors in the 
    simulation expectation for high $\mathcal{F}$ are quite large due to the weighting which
    leaves very few cosmic ray 
    events with high weight in highest $\mathcal{F}$ bins.
  } 
\end{figure}

$\mathcal{F}$ is well-modeled by the simulation as seen in Figure
\ref{datamcbackground}.
Shown is the
experimentally-measured $\mathcal{F}$ distribution for background
cosmic-rays with the simulation expectation overlaid. The
gamma-ray
enhancing event weights from Section \ref{eventweightingsection} 
have been used as a way to probe the data and simulation agreement
under the same conditions as eventual gamma-ray spectral
measurements.
Note that the inclusion of the gamma-ray weights significantly restricts
the number of simulation events surviving to the highest $\mathcal{F}$
bins, where the gamma/hadron separation performs the best. The weights are,
after all, designed to de-emphasize hadronic events.  The expected
background passing rate above $\mathcal{F}$ of 1.6 cannot be reliably estimated 
since no simulated background events survive the weighting.

\subsection{Spectral Fitting}

Since the energy resolution of Milagro is broad, 
typically 50\%-100\%, the energy distribuion expected in 
a $\mathcal{F}$ bin is 
dependent on the spectral 
assumptions. For this reason, we perform all of the spectral fits
in the $\mathcal{F}$ space: 
For each spectral hypothesis, we simulate the expected 
$\mathcal{F}$ distribution and evaluate the goodness of fit based on the 
$\chi^2$ statistic.

We perform spectral fits to a generalized assumption for spectral shape 
described by 
a power law with an exponential cutoff. 
\begin{equation}
  \frac{dN}{dE}(I_0,\alpha,E_{\rm cut}) = I_0 (\frac{E}{E_{0}})^{-\alpha}
  \exp(\frac{-E}{E_{\rm cut}})
\label{fit-function}
\end{equation}
In this equation, $I_0$, $\alpha$ and $E_{cut}$ are fit parameters for the
flux, spectral index and cutoff energy respectively. The $E_{0}$ parameter
is not fit but rather is chosen so that the $\chi^2$ contours of the 
fit variables are 
de-correlated.
This functional form has the benefit that it 
intrinsically models a pure power law hypothesis 
when $E_{cut}$ is above a few hundred TeV and we can test a pure
power law hypothesis 
and a power law with an exponential cutoff hypothesis with one $\chi^{2}$
computation. 

The fit is performed by 
computing $\chi^2$ for a given $\mathcal{F}$ distribution defined as
\begin{equation}
  \chi^2 (I_0,\alpha,E_{\rm cut}) = \sum_{i={
      \mathcal{F} \ bins}} 
  \frac{(P_{i}(I_0,\alpha,E_{\rm cut},{\rm Declination}) - M_{i})^2}
       {\delta P_{i}^2 + \delta M_{i}^2}.
\end{equation}

Here $P$ and $M$ are the sum of the predicted and 
measured sum of weighted events per day from the 
Crab and 
$\delta P$ and $\delta M$ are the error
in $P$ and $M$ respectively. 
We have only considered statistical errors in the
estimation of $\chi^2$.
Notice that since the 
value of $P$ depends on the zenith angle of the source as it transits and 
the distribution
of zenith angles averaged over a transit is the
same for all sources
with the same declination, the predicted daily weight sum depends on the
declination of the source in addition to the hypothesized spectral parameters.

The expected weighted excess is computed for discrete values of 
$\alpha$ between 1.5 and 3.5 in steps of 0.025, $log_{10}(E_{cut})$ 
between 0 and 3 in steps of 0.05. $I_0$ 
was scanned over a range between 0.1 and 4.5 times the nominal pure power 
law Crab
flux measured by HESS \citep{hesscrab} in steps of 0.05.
The values are tabulated 
and
the best fit spectrum is computed by minimizing $\chi^2$.

Section \ref{resultssection} summarizes the results of this technique
applied to the excess from 
the Crab Nebula and to the background cosmic-ray population as a cross-check.


\section {Results}
\label{resultssection}

The success of our technique 
depends on the simulation to 
reliably describe the response of the instrument.
Below 20 TeV, 
the energy spectrum of the Crab has been well measured by 
IACTs.
In 
the range from 20 TeV to 100 TeV the data are limited and somewhat 
contradictory.
We can however 
test
the energy estimation by fitting the spectrum of the hadronic background as a 
cross-check of the method.

\subsection{Systematic Effects}

The 
spectrum of the hadronic background has been well 
measured by a series of balloon-based 
spectrometers 
as well as ground-based air shower detectors. See \cite{pdg} for 
a comprehensive review. 
While the simulation of hadronic interactions introduces
a systematic error that is not present in simulated gamma-ray cascades, 
comparisons with hadronic data are nevertheless
a useful verification of the simulation.
A single day of data is sufficient to fit the cosmic-ray background 
spectrum, so 
reliable and accurate daily fits 
to the cosmic-ray spectrum serve as a measure of the stability of the 
energy response of the instrument.

The hadronic background is composed of numerous species. Protons dominate 
the flux, accounting for about 60\% of the triggers in Milagro, but 
helium ions at about 30\% and heavier element at about 10\% also make 
important contributions. We have simulated the 8 hadronic species with 
the largest contribution: H, He, C, O, Ne, Mg, Si and Fe. The ATIC spectrometer
has measured both the spectrum and the composition of multi-TeV cosmic rays
and found different spectra for the different species. 
We simulate the 8 species 
listed
with their spectra measured by ATIC \citep{aticcr1,aticcr2}
and fit to an overall offset in the spectral index 
($\Delta\alpha$) and a flux scale factor ($S$) where $\Delta\alpha$=0 indicates
that the spectrum was measured to be exactly equal to the ATIC spectrum, 
$\Delta\alpha >$0 indicates a steeper spectrum and $\Delta\alpha < $0 indicates
a flatter spectrum. Similarly, $S$=1 indicates an agreement with the predicted
flux, $S<1$ indicates that Milagro measures a flux that is less than the 
the predicted flux and $S>1$ indicates a greater flux than predicted.

Recall that the eventual gamma-ray fits are performed
using events weighted by the gamma-ray selection weights from 
Section \ref{eventweightingsection}. 
In order that the study of cosmic-ray fits are subjected to the same
systematic effects that may be present in the gamma-ray analysis, 
we use the same event weighting for the cosmic-ray fits, with the 
consequence that the majority of the cosmic-ray events are
given small weight and cosmic-ray showers which appear similar to the
gamma rays receive the most weight.

\begin{figure}
\plottwo{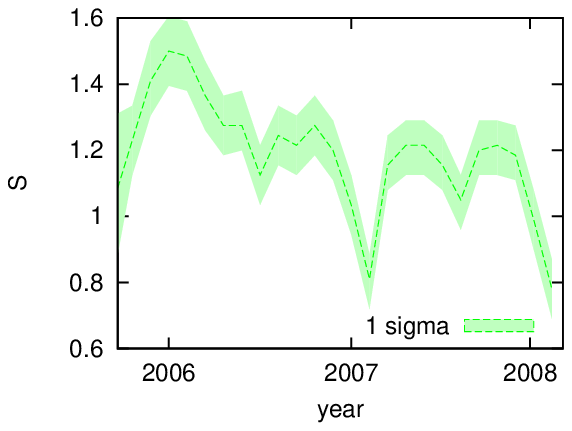}{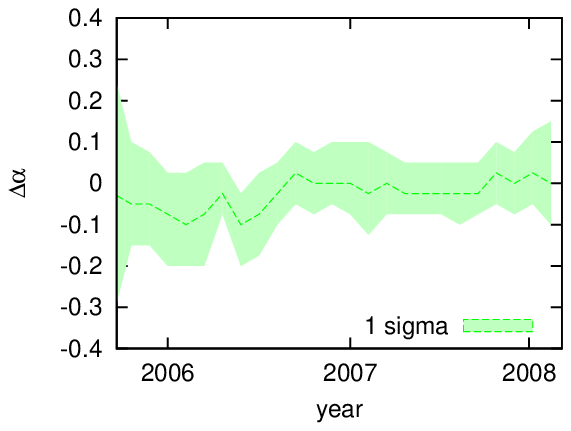}
\caption{The left figure shows the fitted cosmic ray scale (relative to
the simulation) as a function of time and the right figure
shows the fitted cosmic ray index, as an overall steeping or 
flattening of the simulated spectrum.  Note that $\mathcal{F}$
greater than 1.4 was not 
used in these fits.}
\label{fittingcrspectrum}
\end{figure}

Figure \ref{fittingcrspectrum} shows the fit cosmic ray flux 
scaling and spectral index as a function of time. $E_{0}$ was chosen to 
be 10 TeV for these fits.
The cosmic ray index varies by less than 
$\pm0.1$ 
over the 
time shown. The overall flux scaling changes as the operational 
conditions of the
experiment change. Many of these changes
are not included in the simulation. Departures 
from the average are
rare and suggest a systematic uncertainty in the 
total flux of sources close to the \citep{milagrocygnus} of 30\% 
that has been estimated before.

\begin{figure}
\plotone{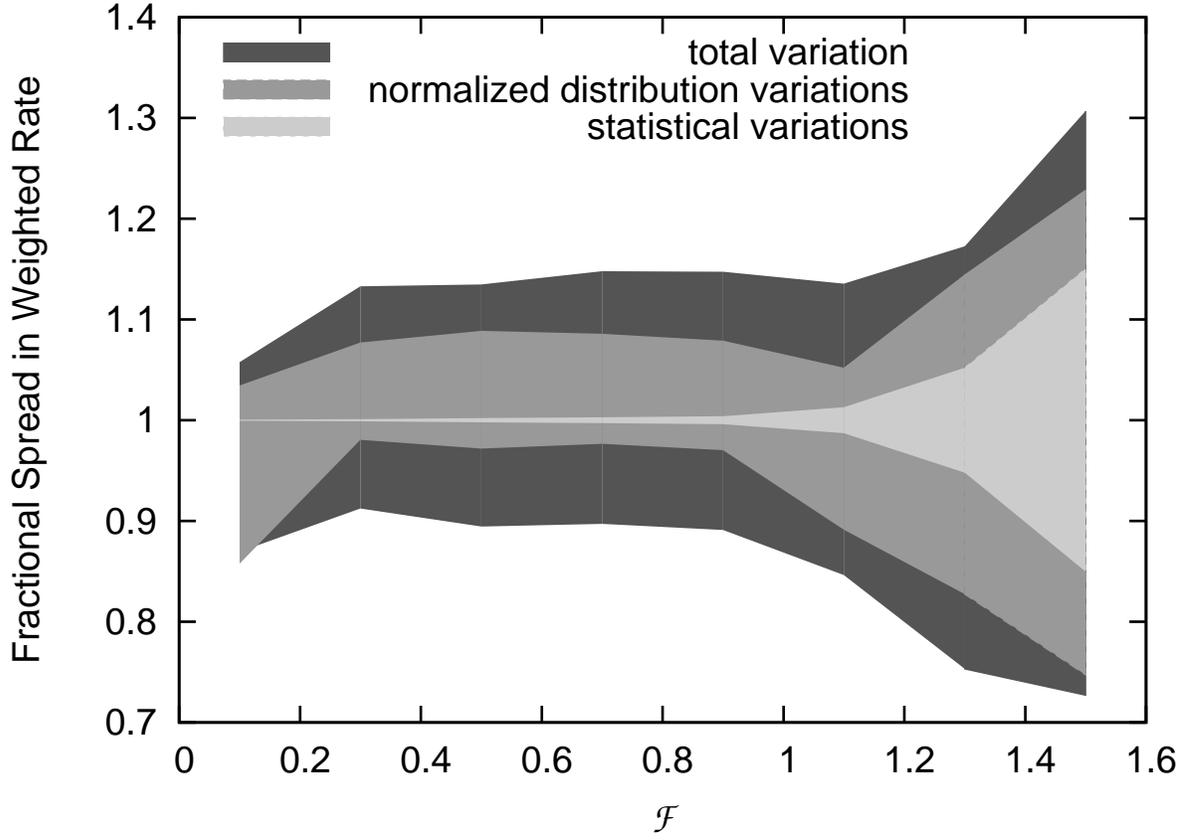}
\caption{This figure quantifies uncertainty in the $\mathcal{F}$ 
distribution. The 68\% central values for background 
weighted event rates going into 
$\mathcal{F}$ bins across different days of data acquisition are shown. 
Both the absolute variations are shown
along with variations after normalizing the underlying $\mathcal{F}$
distributions to unit area. This indicates a fundamental $\sim$ 10\% uncertainty
in the shape of the $\mathcal{F}$ distribution.}
\label{frasorspread}
\end{figure}

The instability of the cosmic-ray fit over time is due to real $\sim 10\%$
changes
in the $\mathcal{F}$ distribution of the data over time. These changes
can be seen easily using the background cosmic-ray data which have
small statistical errors.
Figure \ref{frasorspread} summarizes our uncertainty in the $\mathcal{F}$
distribution based on variations observed in the experimental data.
For each of a set of data runs covering the observation period, we
compute the $\mathcal{F}$ distribution of the background data.
For each bin of $\mathcal{F}$ we quantify the width of the distribution of 
weighted event rates in that bin across the different runs
as the 68\% spread around the median. The fractional
width of these distributions for each $\mathcal{F}$ are shown as the
darkest band in Figure \ref{frasorspread}. Some of the run to run
variation is due to an overall scaling difference between the 
runs. If we normalize the $\mathcal{F}$ of each run to unit area and 
re-do the calculation of the spread across the runs, we get
the darker gray band in Figure \ref{frasorspread}. It is naturally somewhat
smaller than the darkest band because variation that can be attributed
to overall scaling has been taken out. Finally the lightest gray band
shows the fluctuations expected due to purely statistical effects and we can see
that there is a fundamental uncertainty due in the $\mathcal{F}$ distribution
due simply to statistically significant differences in different days of data 
at the level of about 10\%. 
This 
variation is due to real and unmodeled changes in the 
detector calibration, configuration and operating conditions.

\subsection{Spectrum of the Crab}

Applying the method of Section \ref{energyestimationsection} to the statistical
excess from the Crab Nebula, we can determine the spectrum of the Crab Nebula.
The $\chi^{2}$ space that is spanned by the three fitting parameters from 
Equation \ref{fit-function} is scanned to find the global minimum.
To fit a simple power law or to test a number of specific 
hypotheses motivated by measurements from other experiments, one 
of the parameters can be fixed to the assumed value and the minimum
$\chi^{2}$ is then found over the corresponding subset of the fit space.

\begin{figure}
\plotone{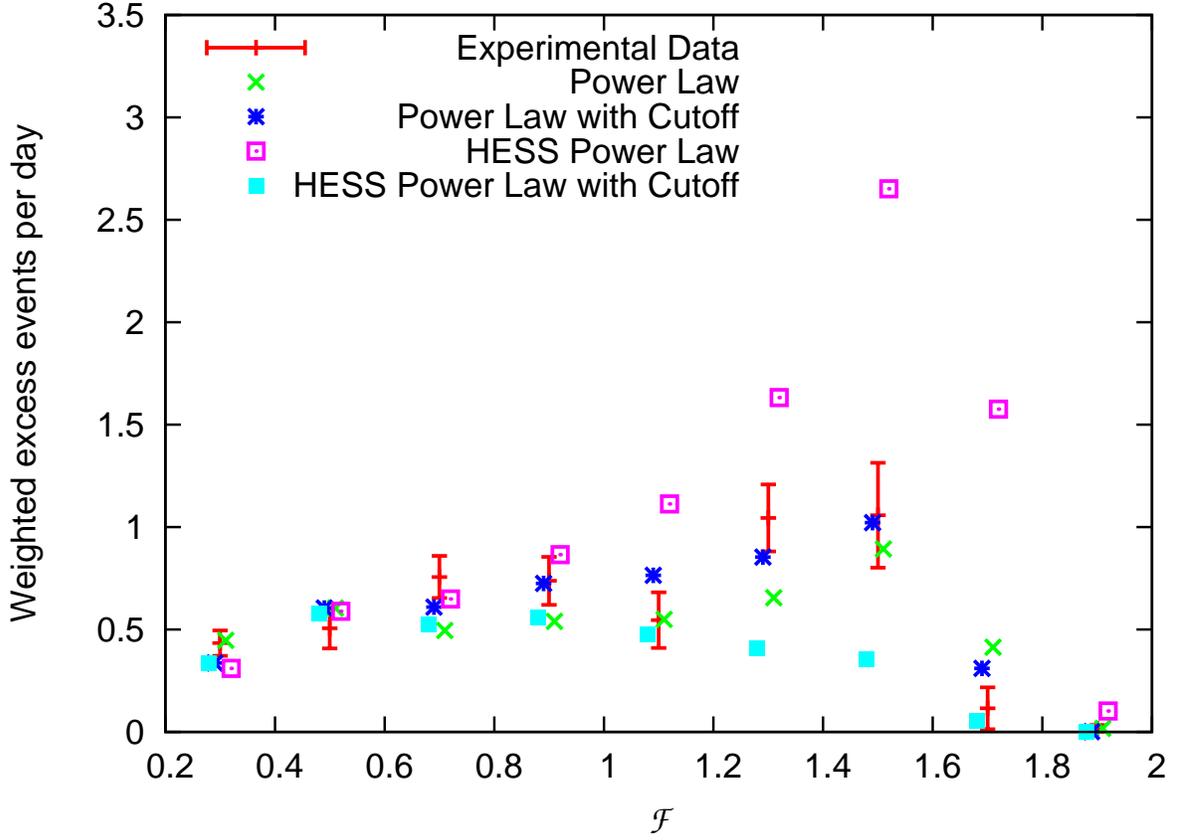}
\caption{
The distribution of the 
$\mathcal{F}$ parameter measured in data and expected from simulation
for several hypotheses. The first two are the best fits to the 
Milagro data both with a pure power-law and 
power-law with an exponential cutoff. The second two show the 
expectations from the best HESS solutions, both for a pure power law (with
a differential photon spectral index of -2.63) and
including an exponential cutoff (with an index of -2.39 and a cutoff at
14.3 TeV) as 
reported in \cite{hesscrab}. Note the points have been offset horizontally
by a small amount for display.
}
\label{data-mc-crab}
\end{figure}

To begin with, we fit the Crab spectrum under the hypothesis that 
the spectrum 
is a pure power law. That is to say that the $E_{\rm cut}$ of Equation
\ref{fit-function} is much higher than the Milagro sensitivity. 
The best fit occurs at $I_o$=
$(6.5\pm0.4_{\rm{(stat)}}) \rm{x} 10^{-14} ({\rm{cm}^2 \cdot \rm{s} \cdot \rm{TeV})^{-1}}$ and $\alpha=3.1\pm0.1_{\rm{(stat)}}$ with $E_{0}$ of 10 TeV.
For this hypothesis, we obtain a $\chi^{2}$ 
of 24.1 with 7 degrees of freedom. The contours of the $\chi^{2}$ function
are ellipsoidal in the space of the fit parameters indicating very little
correlation in the fit parameters. Assuming this hypothesis is right, we
expect to have only a 0.1\% probability to observe a $\chi^{2}$ by chance.
The moderate failure of the two-parameter model to fit the observed
data is robust even if we artificially inflate the error bars in the data
by 10\% (added in quadrature) to allow for our systematic uncertainty in the
rate of events in a given $\mathcal{F}$ bin. With the artificially inflated
error bars, the $\chi^{2}$ improves to 21.7 which corresponds to a 
chance probability
of 0.3\%. Figure \ref{data-mc-crab} shows the $\mathcal{F}$
distribution for the Crab with our best-fit pure power law hypothesis overlaid.

An independent analysis of the Milagro data was done \citep{ballenthesis}
utilizing a different algorithm to estimate the gamma ray energy of each event
which depended on: the core distance of the air shower from the center of the
Milagro pond, the reconstructed zenith angle of the air shower primary, 
and the measured number of PMTs in the top layer and outrigger array. 
Gamma rays were distinguished from cosmic rays using the compactness parameter
\citep{milagrocrab} rather than $A5$. 
The fitted values are consistent 
with the fits obtained with $\mathcal{F}$ and A5 with somewhat larger 
error bars. The agreement
indicates that our reported fit is robust with respect to energy algorithm
and hadron rejection parameter.

\begin{figure}
\plottwo{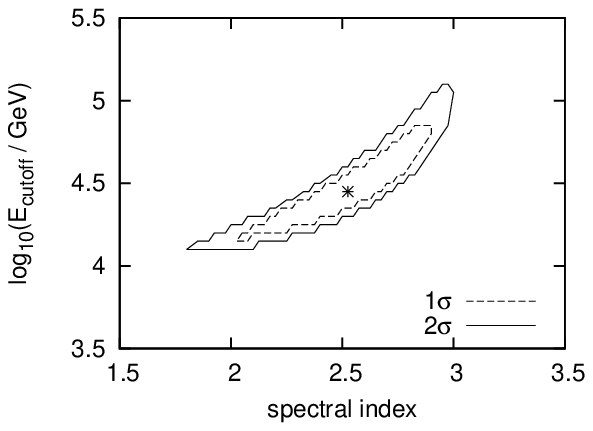}{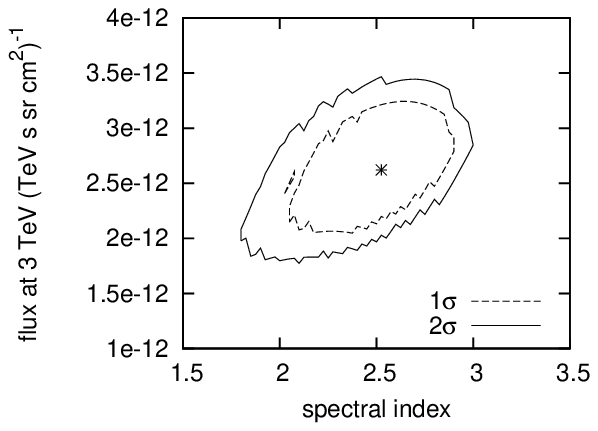}
\caption{Shown are the 1 and 2$\sigma$ allowed regions for the full
power-law hypothesis including an exponential cutoff. The position of 
minimum $\chi^{2}$ is indicated. The 3-d regions have been
projected into planes of the three fit variables.}
\label{crab-chisq}
\end{figure}

We next consider a 
hypothesis of a power law, with an exponential cutoff. This is
Equation \ref{fit-function} where $E_{\rm cut}$ is allowed to vary. With this
additional free parameter, the $\chi^2$ improves to 12.1 with 6 degrees of 
freedom. This corresponds to a chance probability of 6\%.
At the location of the best fit, 
$I_o$=$(2.5^{+0.7}_{-0.4\rm{(stat)}})$ x $10^{-12} ({\rm{cm}^2 \cdot \rm{s} \cdot \rm{TeV})^{-1}}$
with $\alpha=2.5\pm0.4_{\rm{(stat)}}$ and $E_{\rm cut}$=$32^{+39}_{-18\rm{(stat)}}$ TeV. For this 
fit $E_{0}$ was set to 3 TeV.
Figure \ref{crab-chisq} shows projections of the 1 and 2$\sigma$ allowed
regions in the plane of our three fit variables. The 
somewhat broad allowed range of spectral
indices and cutoff energies is due to a fundamental ambiguity in the 
Milagro data that a soft spectrum is hard to distinguish from a harder spectrum
with an exponential cutoff. Fixing the low-energy spectral index 
to the values between 
2.4 to 2.6, as measured by other experiments, gives a 1$\sigma$
allowed range for the cutoff energy of between 20 and 40 TeV.

\begin{figure}
\plottwo{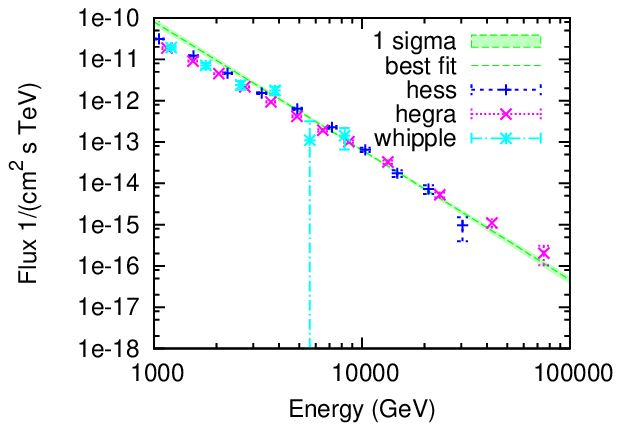}{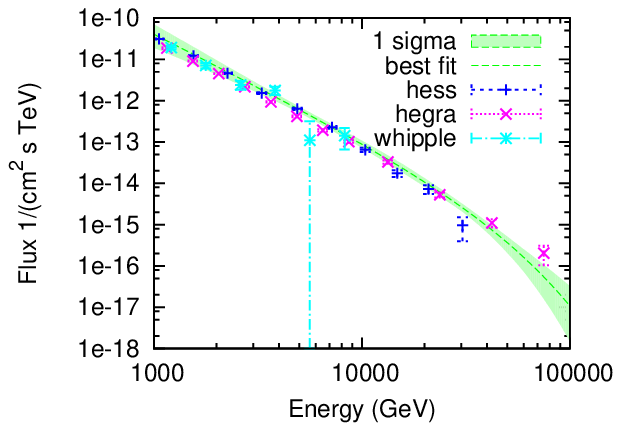}
\caption{
The panels display the spectrum of the Crab as measured by Milagro. The
pure power law hypothesis is shown on the left and the power law with an 
exponential cutoff is shown on the right.
The one
$\sigma$ regions are shown with shading and the points measured
by HESS, HEGRA and Whipple are overlaid.}
\label{crab-fit}
\end{figure}

Neither of the two spectral assumptions is preferred strongly by fitting the
Milagro data.
The pure power law fit is a marginally
poor fit. The addition of an exponential cutoff improves the fit. 
The measured fluxes are shown on Figure \ref{crab-fit} for the two 
hypotheses. Regardless of which fit is chosen, the general conclusion is clear:
The high-energy spectrum, above about 5 - 10 TeV, is steeper than measured by
IACTs at lower 
energies. In the pure power-law hypothesis, this manifests itself
as a measured spectrum of $\alpha=3.1\pm0.1$, steeper than has been 
measured by,
for instance, HESS of 2.4 to 2.6. The fit that allows for an exponential
cutoff reproduces the low-energy spectral index measured by IACTs and this
steepening at high energy is seen as an exponential cutoff at $\sim$ 30 TeV.

Finally, it is interesting to note that above 30 TeV, the HESS and HEGRA
data are mildly inconsistent. The HEGRA measurement 
continues to higher
energy than the HESS data. It has been suggested \citep{crabinterp} that
this discrepancy is related to the time variability observed by the Crab
since HEGRA was observing earlier than HESS.
The Milagro data, which
represents the time-average over 3 years of data, indicates
a spectrum between 
the data of HESS and HEGRA.

\section{Conclusions}

The Crab Nebula is the brightest northern hemisphere TeV source and
has been extensively measured by IACTs above 1 TeV.
The Milagro measurement of the energy spectrum of the Crab has been 
presented.
A background rejection parameter ($A5$)
has been described and shown to distinguish between
gamma ray and hadronic primaries in the detector. 
We have presented the weighting and background
estimation and background subtraction techniques used to extract the 
Crab signal, giving a
17$\sigma$ over the lifetime of the experiment. 

The size of an air shower at ground represented by the fraction of 
PMTs in the Milagro experiment that detect a signal 
(the $\mathcal{F}$ parameter),
is a suitable variable for measuring the spectra of primary 
TeV gamma and cosmic rays.
The
relatively simple form of $\mathcal{F}$ is justified because the
dominant effect contributing to the energy resolution of Milagro is 
fluctuations in the depth of first interaction of the primary
particles and not in the measurement of the energy reaching
the ground.  The parameter is well modeled in the simulation as
observed by studying the cosmic ray background. 

The energy spectrum of gamma rays from the Crab 
between 1 and 100 TeV has been
measured
by fitting the observed $\mathcal{F}$ distribution of the
Crab with expectations from simulation. 
A steepening of the spectrum above about 5-10 TeV with respect to 
measurements by IACTs at lower energies has been measured.

The experiment observes the entire overhead sky, the data and
analysis technique presented here 
for the Crab observations can be used to measure the flux and spectral
properties of the other sources in the Milagro catalog.  The agreement
seen on the
Crab as a calibration source justifies confidence in measurements
of other sources.

\acknowledgments

We gratefully acknowledge Scott Delay and Michael Schneider for their
dedicated efforts in the construction and maintenance of the Milagro
experiment. This work has been supported by the National Science Foundation
(under grants PHY-0245234, -0302000, -0400424, -0504201, -0601080,
and ATM-0002744), the US Department of Energy (Office of High-Energy Physics
and Office of Nuclear Physics), Los Alamos National Laboratory, the
University of California, and the Institute of Geophysics and Planetary
Physics.

\bibliography{bibliography}

\label{cosrayspec}

\end{document}